\documentclass[12pt]{article}
\usepackage{amsmath,amssymb}
\usepackage{graphicx}
\usepackage{color}
\begin{document}
\baselineskip=16pt
\begin{titlepage}
\begin{flushright}
{\small SU-HET-04-2014}
\end{flushright}
\vspace*{1.2cm}

\begin{center}

{\Large\bf Higgs inflation with singlet scalar dark matter and right-handed neutrino in light of BICEP2}
\lineskip .75em
\vskip 1.5cm

\normalsize
{\large Naoyuki Haba} and {\large Ryo Takahashi}

\vspace{1cm}
{\it Graduate School of Science and Engineering, Shimane University, 

Matsue, Shimane 690-8504, Japan}

\vspace*{10mm}

{\bf Abstract}\\[5mm]
{\parbox{13cm}{\hspace{5mm}
We discuss the Higgs inflation scenario with singlet scalar dark matter and a 
right-handed neutrino. The singlet scalar and the right-handed neutrino play crucial roles for 
realizing a suitable plateau of Higgs potential with the center value of the top
 mass of Tevatron and LHC measurements. This Higgs inflation scenario predicts 
about a 1 TeV scalar dark matter and an $\mathcal{O}(10^{14})$ GeV right-handed neutrino 
by use of 125.6 GeV Higgs mass, 173.34 GeV top mass, and a nonminimal gravity 
coupling $\xi\simeq10.1$. This inflation model is consistent with the recent result of the tensor-to-scalar 
ratio $r=0.20_{-0.05}^{+0.07}$ by the BICEP2 Collaboration.}}

\end{center}
\end{titlepage}

\section{Introduction}

The Higgs particle has been discovered at the CERN Large Hadron Collider 
experiment, and their results are almost consistent with the standard model 
(SM)~\cite{Chatrchyan:2013lba,CMS}. In addition, since the experiment has not 
obtained evidence of new physics so far (e.g., supersymmetry or extra dimension(s), 
etc.), one might consider a scenario such that the SM is valid up to a 
very high energy scale (GUT, string, or Planck scales). In fact, there have been  
several curious research results for this scenario. For example, Ref.~\cite{Froggatt:1995rt}
 showed that the multiple point criticality principle\footnote{The principle 
says that there are two degenerate vacua in the Higgs potential of SM. One is at
 the Planck scale and another one is at the electroweak (EW) scale where we 
live.} predicts $135\pm9$ GeV Higgs and $173\pm5$ GeV top masses. 
Reference~\cite{Shaposhnikov:2009pv} also pointed out that 126 GeV Higgs mass 
can be realized with a few GeV uncertainty in an asymptotic safety scenario of 
gravity. They clarified that the vanishing Higgs self-coupling and its 
$\beta$-function at the Planck scale, $\lambda(M_{\rm pl})=\beta_\lambda(M_{\rm 
pl})=0$, lead to the above values of the Higgs and top masses, which are close 
to the current experimental values. Reference~\cite{Holthausen:2011aa} 
investigated the realization of the Veltman condition, Str$M^2(M_{\rm pl})=0$, and
 the vanishing anomalous dimension of the Higgs mass, $\gamma_{m_h}(M_{\rm pl})=0$,
 at the Planck scale in addition to $\lambda(M_{\rm pl})=\beta_\lambda(M_{\rm 
pl})=0$. As a result, the authors could find that the realization of the BCs 
predicts 127-142 GeV Higgs mass. It is interesting that the above BCs can lead 
to close values of the Higgs and top masses to the experimental ones, but it 
seems difficult to reproduce the experimental center values of the top and Higgs
 masses at the same time (see also 
Refs.~\cite{Degrassi:2012ry}-\cite{Masina:2013wja} for the recent analyses). 
Since the realization of BCs means that the Higgs potential is almost flat near 
the Planck scale, an application of the flat potential to the inflation, the 
so-called Higgs inflation~\cite{Bezrukov:2007ep}-\cite{Hamada:2013mya}, is 
intriguing. In addition, recently, the tensor-to-scalar ratio,
 \begin{eqnarray}
  r=0.20_{-0.05}^{+0.07},
 \end{eqnarray}
was reported by the BICEP2 Collaboration~\cite{Ade:2014xna}, and several 
researches have investigated in the ordinary Higgs inflation and models related 
with the Higgs field~\cite{Nakayama:2014koa}-\cite{Feng:2014naa}. In particular,
 the authors of Ref.~\cite{Hamada:2014iga} pointed out that the Higgs potential 
with the small top mass and a nonminimal coupling $\xi=7$ can make the ordinary
 Higgs inflation consistent with the BICEP2 result. 

In this paper, we will investigate the Higgs inflation with the singlets 
extension of the SM. The gauge singlet fields can play various roles in 
models/theories beyond the SM. For instance, a singlet real scalar field can 
rescue the SM from the vacuum instability, and it can be a candidate for dark 
matter (DM) with odd parity under an additional $Z_2$ symmetry (e.g., 
see~\cite{Silveira:1985rk}-\cite{Boucenna:2014uma}). In addition, a scalar can 
play an important role of EW and conformal symmetry breaking through a strongly 
coupled hidden sector (see~\cite{Hur:2007uz,Hur:2011sv,Holthausen:2013ota} for 
more recent discussion). It is also well known that the right-handed neutrinos 
can generate tiny active neutrino masses through a seesaw mechanism and the 
baryon asymmetry of the Universe (BAU) through the leptogenesis.

The singlet scalar and the right-handed neutrino play crucial roles for 
realizing a suitable plateau of Higgs potential with the center value of the top
 mass of Tevatron and LHC measurements~\cite{ATLAS:2014wva}. We will show that 
this Higgs inflation scenario predicts about a 1 TeV scalar DM and an  
$\mathcal{O}(10^{14})$ GeV right-handed neutrino by use of a 125.6 GeV Higgs 
mass, a 173.34 GeV top mass, and a nonminimal gravity coupling $\xi\simeq10.1$.
 We stress that the inflation model is consistent with the recent result of the 
tensor-to-scalar ratio $r=0.20_{-0.05}^{+0.07}$ by the BICEP2 Collaboration.

\section{Singlets extension of the SM}

We discuss the SM with a real singlet scalar and a right-handed neutrino. The 
relevant Lagrangians of the model are given by
 \begin{eqnarray}
  \mathcal{L}        
   &=      & \mathcal{L}_{\rm SM}+\mathcal{L}_S+\mathcal{L}_N, \\
  \mathcal{L}_{\rm SM} 
   &\supset& -\lambda\left(|H|^2-\frac{v^2}{2}\right)^2, \label{S}\\
  \mathcal{L}_S      
   &=      & -\frac{\bar{m}_S^2}{2}S^2-\frac{k}{2}|H|^2S^2-\frac{\lambda_S}{4!}S^4
             +({\rm kinetic~term}), \\
  \mathcal{L}_N
   &=      & -\left(\frac{M_R}{2}\overline{N^c}N
                    +y_N\overline{N}L\tilde{H}+c.c.\right)
             +({\rm kinetic~term}),
 \end{eqnarray}
with $\tilde{H}=-i\sigma_2H^\ast$, where $\mathcal{L}_{\rm SM}$ is the SM 
Lagrangian including the Higgs potential. $H$ is the Higgs doublet, $v$ is the 
vacuum expectation value of the Higgs, $L$ is the left-handed lepton doublet in
 the SM, $S$ is a gauge singlet real scalar, and $N$ is a right-handed neutrino.
 We omit the flavor index of left-handed lepton doublets and assume that only 
the singlet real scalar has odd parity under an additional $Z_2$. Thus, the 
singlet scalar can be DM when appropriate its mass and coupling $k$ are taken.
 The right-handed neutrino can generate the tiny neutrino mass through the
 type-I seesaw mechanism.

The RGEs of $(\lambda,k,\lambda_S)$ are given by
 \begin{eqnarray}
  (4\pi)^2\frac{dX}{dt}=\beta_X~~~(X=\lambda,k,\lambda_S),
 \end{eqnarray}
with
 \begin{eqnarray}
  \beta_\lambda
   &=& 24\lambda^2+4\lambda(3y^2+y_N^2)-2(3y^4+y_N^4)-3\lambda(g'{}^2+3g^2)
       +\frac{3}{8}\left[2g^4+(g'{}^2+g^2)^2\right]+\frac{k^2}{2}, \label{l} \\
  \beta_k
   &=& k\left[4k+12\lambda+2(3y^2+y_N^2)-\frac{3}{2}(g'{}^2+3g^2)+\lambda_S\right], \label{k} \\
  \beta_{\lambda_S}
   &=& 3\lambda_S^2+12k^2,
 \end{eqnarray}
at the one-loop level, where $y$ ($y_N$) is the top (neutrino) Yukawa coupling, 
$g$ and $g'$ are gauge couplings, $t$ is defined as $t\equiv\ln(\mu/1~{\rm 
GeV})$, $\mu$ is a renormalization scale within $M_Z\leq\mu\leq M_{\rm pl}$, $M_Z$
 is the $Z$ boson mass, and $M_{\rm pl}$ is the reduced Planck mass as $M_{\rm 
pl}=2.435\times10^{18}$ GeV. When $\mu$ is smaller than a mass of particle, 
contribution to the $\beta$-functions from the particle should be subtracted. 
For example, the terms proportional to $y_N$ in Eqs.~(\ref{l}) and (\ref{k}) 
disappear in an energy range of $\mu<M_R$. Typical properties of evolutions of scalar quartic 
couplings are listed as follows:
\begin{itemize}
\item An evolution of $k$ is small when $k(M_Z)$ is small, because 
$\beta_k$ is proportional to $k$ itself. In this case, the evolution of 
$\lambda(\mu)$ resembles that of the SM, closely.

\item When one takes the experimental center values of the Higgs and top masses,
 $\lambda(\mu)$ is negative within a region of $\mathcal{O}(10^{10})~{\rm 
GeV}\lesssim\mu\leq M_{\rm pl}$ (see the dotted curve in Fig.~\ref{fig1} (a)). It 
is known as the vacuum instability or metastability. This is caused by the 
negative contribution proportional to the top Yukawa coupling $-6y^4$ to 
$\beta_\lambda$ in Eq.~(\ref{l}). There exists a minimum in the evolution of 
$\lambda(\mu)$ around $\mu\sim\mathcal{O}(10^{17})$ GeV. But, for taking a 
heavier Higgs mass as $127~{\rm GeV}\lesssim m_H\lesssim130~{\rm GeV}$ with 
$M_t=173.1\pm0.6~{\rm GeV}$ or a lighter top mass as $171.3~{\rm GeV}\lesssim 
M_t\lesssim171.7~{\rm GeV}$ with $m_H=126~{\rm GeV}$, $\lambda(\mu)$ can be 
positive over a region of $M_Z\leq\mu\leq M_{\rm pl}$ in NNLO 
calculations~\cite{Degrassi:2012ry}.

\item The additional term 
$+k^2/2$ contributes to $\beta_\lambda$, which can lift the evolution of 
$\lambda(\mu)$ not to be negative up to the Planck scale (see the dashed curve in 
Fig.~\ref{fig1} (a)). On the other hand, the contribution from the 
Yukawa coupling $-2y_N^4$ pushes down the evolution of 
$\lambda(\mu)$ like the top Yukawa coupling (compare the dashed curve with the solid one
 in Fig.~\ref{fig1} (a)). The value of $\mu_{\rm min}$, where $\lambda_{\rm 
min}\equiv\lambda(\mu_{\rm min})={\rm min}\{\lambda(\mu)\}$, shifts smaller (larger)
 than $\mathcal{O}(10^{17})$ GeV by introducing $S$ ($N$) 
because the positive (negative) term $+k^2/2$ ($-2y_N^4$) contributes to 
$\beta_\lambda$. These features will be crucial in our realization of the successful Higgs 
inflation in singlets extension of the SM; i.e. we will fine-tune between these two contributions to obtain the suitable plateau in the Higgs inflation potential which is consistent
with the recent BICEP2 result within the experimental range as
$M_t=173.34\pm0.76$ GeV~\cite{ATLAS:2014wva}.

\item The evolution of $\lambda_S(\mu)$ is a monotonical increasing function of 
the renormalization scale, and $\lambda_S(\mu)$ does not contribute to $\beta_\lambda$
 directly. 
\end{itemize}
Let us show that the Higgs inflation works well in this model as below.

\section{Higgs inflation in singlets extension of the SM}

We start with the relevant action of the ordinary Higgs 
inflation~\cite{Bezrukov:2007ep} as
 \begin{eqnarray}
  S_J\supset\int d^4x\sqrt{-g}\left(-\frac{M_{\rm pl}^2+\xi h^2}{2}R
                                   +\mathcal{L}_{\rm SM}\right),
 \end{eqnarray}
in the Jordan frame, where $\xi$ is the nonminimal coupling to the Ricci scalar
 $R$, $H=(0,h)^T/\sqrt{2}$ is taken in the unitary gauge, and $\mathcal{L}_{\rm 
SM}$ includes the Higgs potential given in Eq.~(\ref{S}). After the conformal 
transformation from the Jordan frame to the Einstein one 
($\hat{g}_{\mu\nu}=\Omega^2g_{\mu\nu}$ and $\Omega^2\equiv1+\xi h^2/M_{\rm pl}^2$), one
 can write down the relevant action as
 \begin{eqnarray}
  S_E\supset\int d^4x\sqrt{-\hat{g}}
                \left(-\frac{M_{\rm pl}^2}{2}\hat{R}
                      +\frac{\partial_\mu\chi\partial^\mu\chi}{2}
                      -\frac{\lambda}{4\Omega(\chi)^4}(h(\chi)^2-v^2)^2\right),
 \end{eqnarray}
where $\hat{R}$ is given by $R$ and $\hat{g}_{\mu\nu}$, and $\chi$ is a 
canonically normalized field as
 \begin{eqnarray}
  \frac{d\chi}{dh}=\sqrt{\frac{\Omega^2+6\xi^2h^2/M_{\rm pl}^2}{\Omega^4}}.
 \end{eqnarray}
The slow roll parameters for the inflation are calculated as
 \begin{eqnarray}
  \epsilon=\frac{M_{\rm pl}^2}{2}\left(\frac{dU/d\chi}{U}\right)^2,~~~
  \eta=M_{\rm pl}^2\frac{d^2U/d\chi^2}{U},
 \end{eqnarray}
with
 \begin{eqnarray}
  U(\chi)\equiv\frac{\lambda}{4\Omega(\chi)^4}(h(\chi)^2-v^2)^2.
 \end{eqnarray}
Then, the spectral index and the tensor-to-scalar ratio are given by 
$n_s=1-6\epsilon+2\eta$ and $r=16\epsilon$, respectively. The number of 
$e$-foldings is
 \begin{eqnarray}
  N=\int_{h_{\rm end}}^{h_0}\frac{1}{M_{\rm pl}^2}\frac{U}{dU/dh}
    \left(\frac{d\chi}{dh}\right)^2dh,
 \end{eqnarray}
where $h_0$ ($h_{\rm end}$) is the initial (final) value when the inflation starts
 (ends). $h_{\rm end}$ is given as the slow roll conditions 
($\epsilon,|\eta|\ll1$) are broken.

It is known that the SM Higgs potential can have a plateau by taking a 
fine-tuned {\it small} top mass of $M_t=171.0789~(171.0578)$ GeV for $m_H=125.6~(125)$
 GeV~\cite{Hamada:2013mya,Hamada:2014iga,Fairbairn:2014nxa}. By using this 
plateau, the authors of Ref.~\cite{Hamada:2014iga} pointed out that $r\simeq0.2$ can be 
achieved by introducing $\xi=7$ in the Higgs inflation. On the other hand, if 
the plateau in not used, the value of $\xi$ should be as large as 
$\xi\sim\mathcal{O}(10^4)$ in order to have enough $e$-foldings. However, in this case, $r$ becomes too tiny as $r\simeq3.3\times10^{-3}$ to be consistent with the recent BICEP2 result, since the potential is too flat at the beginning of the inflation. The Higgs inflation with the plateau induced from 
$M_t\simeq171.1$ GeV and $\xi=7$ does not suffer from this problem, since a 
suitable $e$-foldings ($50\lesssim N\lesssim60$) and $r\simeq0.2$ are realized 
at the same time. But, this top mass is out of $M_t=173.34\pm0.76$ 
GeV~\cite{ATLAS:2014wva}, anyhow.

Now let us try to obtain a suitable Higgs inflation to realize $m_H\simeq125.6$ 
GeV, $M_t\simeq173.34$ GeV, $r\simeq0.2$, and $50\lesssim N\lesssim60$ as well as suitable DM
 and neutrino masses in the singlet extension of the SM. The realization of the 
scenario can be understood by investigating the behavior of $\lambda(\mu)$. A 
typical evolution of $\lambda(\mu)$ in the model is shown in 
Fig.~\ref{fig1} (a):
\begin{figure}
\hspace{4.9cm}(a)\hspace{7.3cm}(b)
\begin{center}
\includegraphics[scale=0.8]{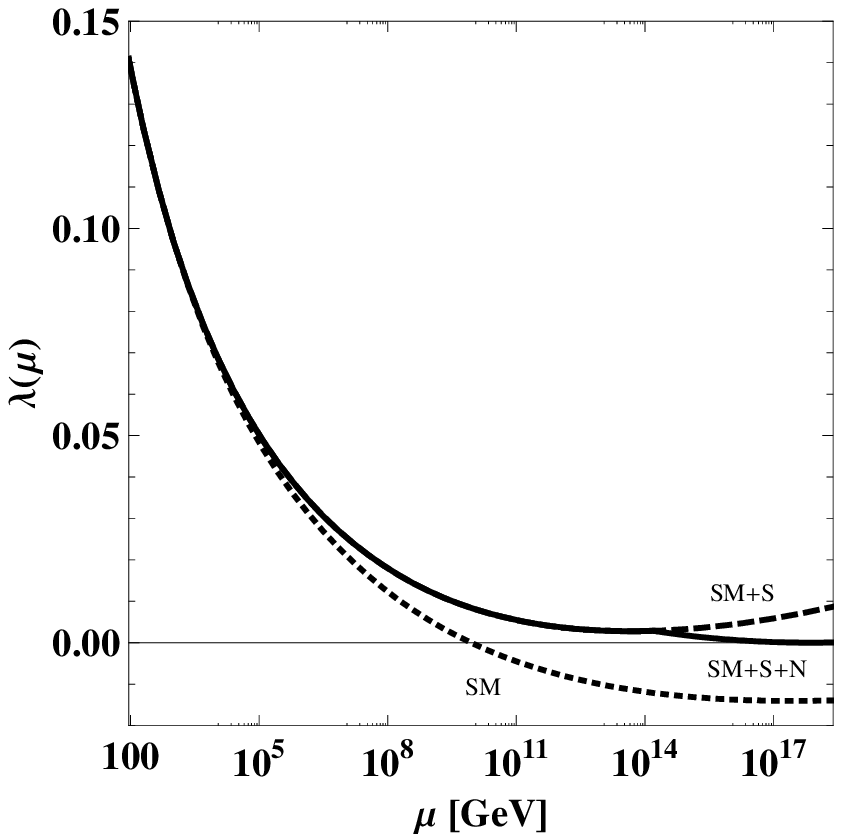}\hspace{5mm}
\raisebox{0.5mm}{\includegraphics[scale=0.69]{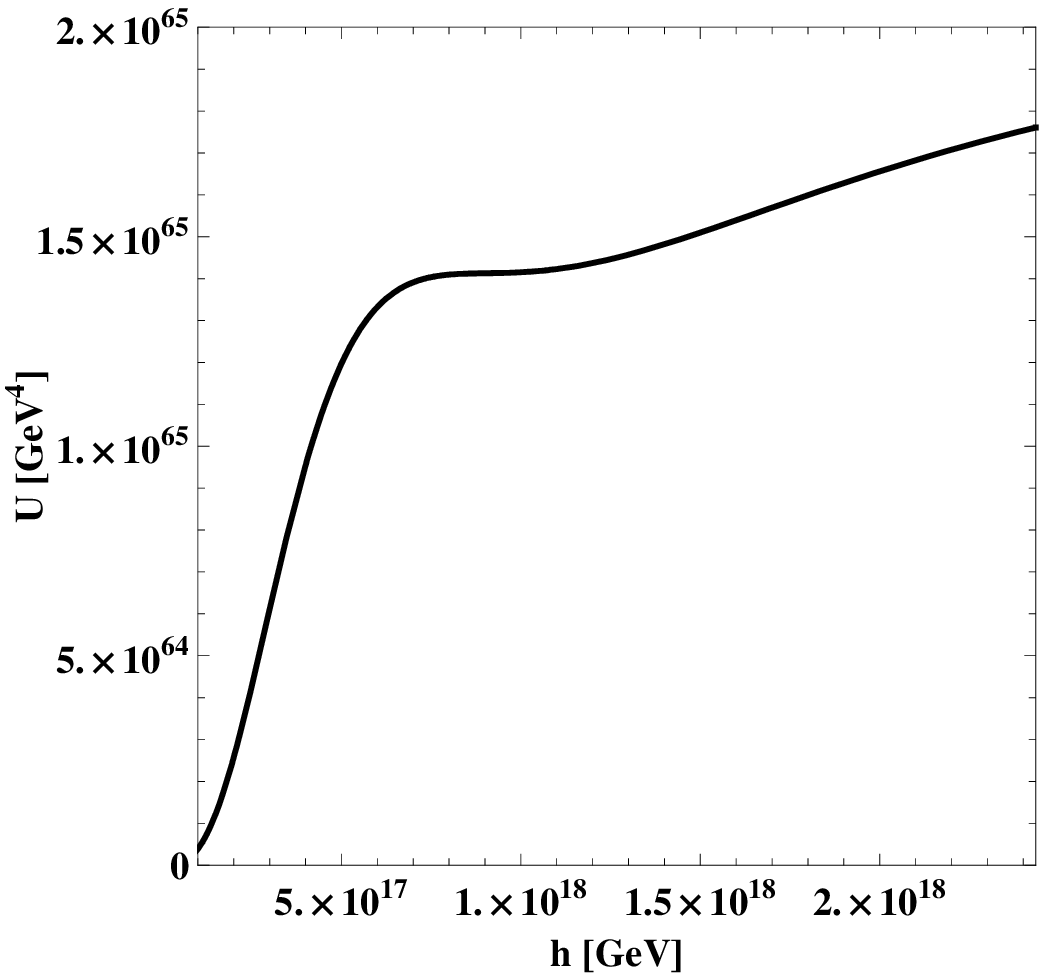}}
\end{center}
\caption{(a) A typical evolution of $\lambda(\mu)$, and (b) the scalar potential
 in the singlets extension of the SM are shown. We take $M_t=173.34$ GeV, 
$m_H=125.6$ GeV, $m_S\simeq1029$ GeV, $M_R\simeq1.58\times10^{14}$ GeV, 
$k(M_Z)\simeq0.325$, and $y_N(M_Z)=\sqrt{m_\nu M_R}/v\simeq0.512$. In (a), 
dotted, dashed and solid curves indicate typical evolutions of $\lambda(\mu)$ in
 the SM, SM with a singlet scalar (SM+$S$), and SM with a single scalar and a 
right-handed neutrino (SM+$S$+$N$), respectively.}
\label{fig1}
\end{figure}\begin{itemize}
\item At first, $\lambda(\mu)$ in the SM is depicted by the dotted curve when $M_t=173.34$ GeV 
and $m_H=125.6$ GeV.

\item Next, we add $S$ with mass $m_S\simeq1029$ GeV and 
coupling $k(M_Z)\simeq0.325$ into the SM. $\lambda(\mu)$ in the case is shown by
 dashed curve. It is seen that the model can avoid the vacuum instability, but 
the value of $\mu_{\rm min}$ becomes smaller than that of the SM. This is 
problematic for the inflation because one cannot have a plateau around 
$\mu\sim\mathcal{O}(10^{17-18})$ GeV.

\item Next is the case of introducing a heavy right-handed neutrino of $M_R\sim\mathcal{O}(10^{14})$ GeV with a suitable $y_N$, where the evolution of $\lambda(\mu)$ is pushed down again. Then, $\mu_{\rm min}\sim\mathcal{O}(10^{17-18})$ 
GeV and $10^{-6}<\lambda(\mu_{\rm min})\lesssim10^{-5}$ can be realized by a 
fine-tuning of $M_R$. In Fig.~\ref{fig1} (a) and (b), values of $M_R$ and  $y_N(M_Z)$ are taken to reproduce a typical active neutrino mass of $m_\nu=0.1$ eV. 
\end{itemize}
The resultant scalar potential for the inflation is shown in Fig~\ref{fig1} (b). Stress that the experimental center value of the top mass $M_t=173.34$ GeV can be used due to the effects of $S$ and $N$.

Finally, we show explicit magnitudes of all parameters which realize the successful Higgs inflation. They are 
 \begin{eqnarray}
  &&m_S\simeq1029.492~{\rm GeV},~~~M_R\simeq1.583687\times10^{14}~{\rm GeV},~~~
    m_\nu=0.1~{\rm eV}, \nonumber \\
  &&k(M_Z)\simeq0.3249353,~~~\lambda_S(M_Z)=0.1,~~~\xi|_{\mu=h_0}=10.097, 
    \nonumber
 \end{eqnarray}
with the experimental center values of
 \begin{eqnarray}
  m_H=125.6~{\rm GeV},~~~M_t=173.34~{\rm GeV},~~~\alpha_s(M_Z)^{-1}=0.1184. \nonumber
 \end{eqnarray}
They reproduce\footnote{Here we include 2-loop SM contributions to $\beta_\lambda$. }
 \begin{eqnarray}
  r\simeq0.200,~~~n_s\simeq0.955,~~~N\simeq50.6. \nonumber
 \end{eqnarray}
The value of $k(M_Z)$ is 
determined by the condition that the $S$ can account for the relic abundance of DM (e.g. see~\cite{Cline:2013gha,HKT}), i.e. 
$\Omega_S\bar{h}^2=0.119$ where $\Omega_S$ and $\bar{h}$ are the density 
parameter of the singlet scalar DM and the Hubble constant, respectively. The 
value of $\lambda_S(M_Z)$ is irrelevant to our result, as long as 
 of $0\leq\lambda_S(M_Z)<1$ as discussed in~\cite{HKT}. The value of $y_N$ is determined by the seesaw formula, $m_\nu=(y_Nv)^2/M_R$ with $m_\nu=0.1$ eV and 
$v=246$ GeV, where the right-handed neutrino can generate one active neutrino mass. Other neutrino masses can be realized by introducing lighter right-handed 
neutrinos with smaller neutrino Yukawa couplings. It is because the neutrino 
Yukawa couplings do not affect on the RGE evolution when they are smaller than 
bottom Yukawa coupling. With the above conditions, the values of $(m_S,M_R,\xi)$
 are uniquely determined to achieve realistic magnitudes of the cosmological 
parameters $(r,n_s,N)$ under given values of $(m_H,M_t,m_\nu,\lambda_S,\alpha_S)$.
 Our solution also indicates
 \begin{eqnarray}
  &&\mu_{\rm min}\simeq 7.50\times10^{17}~{\rm GeV},~~~\lambda_{\rm min}\simeq2.44\times10^{-6}, \nonumber \\
  &&h_0\simeq1.86\times10^{18}~{\rm GeV},~~~h_{\rm end}\simeq4.61\times10^{17}~{\rm GeV},~~~U(h_0)=1.6\times10^{65}~{\rm GeV}^4, \nonumber 
 \end{eqnarray}
which realize the successful Higgs inflation. If one considers slightly lighter 
(heavier) DM mass, $\mu_{\rm min}$ or $\lambda_{\rm min}$ becomes too small (large) 
to achieve a realistic inflation even by fine-tuning of $M_R$ and $\xi$. This 
model has 1029 GeV DM mass, which is consistent with DM 
experiments~\cite{Akerib:2013tjd} (see also~\cite{Cline:2013gha}). It might be 
detected by the future experiment such as XENON1T, XENON100 with 20 times 
sensitivity, combined analysis of Fermi+CTA+Planck observations, 
etc.~\cite{Cline:2013gha}. When we take care of experimental uncertainties 
(and a tiny effect from $\lambda_S$), we can draw allowed regions around the
 typical point shown above. 
 
\section{Summary}
We have investigated the Higgs inflation scenario with singlet scalar dark matter and the right- handed neutrino. The singlet scalar and the right-handed neutrino play crucial roles for realizing the suitable plateau of Higgs potential with the center value of the top mass of Tevatron and LHC measurements. We have shown that this Higgs inflation scenario predicts a 1029 GeV scalar DM and an $\mathcal{O}(10^{14})$ GeV right-handed neutrino by use of 125.6 GeV
Higgs mass, 173.34 GeV top mass, and a nonminimal gravity coupling $\xi\simeq10.1$. This inflation model works well completely, and it is consistent with the recent result of
tensor-to-scalar ratio $r=0.20_{-0.05}^{+0.07}$ by the BICEP2 Collaboration.

\subsection*{Acknowledgement}

The authors thank Kunio Kaneta and Hiroyuki Ishida for fruitful discussions at an early stage of 
this work and pointing out typos in the manuscript. This work is partially supported by Scientific Grant by Ministry of 
Education and Science, No. 24540272. The work of R.T. is supported by Research Fellowships of the Japan Society for the Promotion of Science for Young Scientists.

\end{document}